\documentclass[prd,preprint,superscriptaddress,showpacs]{revtex4}

\usepackage{revsymb}
\usepackage{graphicx}

\newcommand{\be}{\begin{equation}}
\newcommand{\ee}{\end{equation}}
\newcommand{\ba}{\begin{eqnarray}}
\newcommand{\ea}{\end{eqnarray}}
\newcommand{\no}{\noindent}
\newcommand{\n}{\label}

\begin{document}

\title{Internal symmetry in Bianchi type-I cosmologies}

\author{Luis P. Chimento}
\altaffiliation[]
{Fellow of the Consejo Nacional de Investigaciones
Cient\'{\i}ficas y T\'{e}cnicas.}
\email{chimento@df.uba.ar}
\affiliation
{Departamento de F\'{\i}sica,
Facultad de Ciencias Exactas y Naturales,
Universidad de Buenos Aires,
Ciudad  Universitaria,  Pabell\'on  I,
1428 Buenos Aires, Argentina.}[]

\bibliographystyle{plain}

\begin{abstract}

Using the one-parameter internal symmetry group in the Bianchi type-I
spacetime for cosmological models with a perfect fluid, we show that a system
of coordinates exists in the associated internal space where two scale factors
become equal. We find the general solution for an anisotropic model containing
a perfect fluid with constant baryotropic index and investigate the asymptotic
regimes. We obtain exact solutions for a set of anisotropic fluids which
includes an anisotropic stiff fluid.

\end{abstract}

\vskip 1cm


\maketitle

\newpage

\section{Introduction}

Einstein's classical theory of gravitation has been successfully applied to
the solution of problems concerning astronomical scales of length and time,
such as the formulation of cosmological models. In the development of
cosmology much effort is being focused on models in which the universe was
expanding at very early stages of its evolution. There is observational
evidence that at present the Universe seems homogeneous and has been highly
isotropic since the recombination era. In fact, observations of the cosmic
microwave background radiation, galaxies, and other astronomical objects
reveal that our universe, on a very large scale, is remarkably uniform
\cite{cobe,barrow} and is currently under accelerated expansion
\cite{nova}-\cite{tur}. It follows that on large scales the accelerated
expansion is the mechanism to solve the problem of the apparent large-scale
isotropy of the universe and its behaviour is approximately described by a
Friedmann-Robertson-Walker (FRW) model.

The claimed isotropy of the universe is deduced from reliable measurements,
which are the most accurate ones in cosmology. However, these results have
been established for times after the era at which the universe became
transparent to radiation. Their extrapolation to earlier times and, in
particular, near the Planck or string scale is totally unfounded. There are
theoretical arguments that sustain the existence of an anisotropic phase that
approaches an isotropic one \cite{Mis}. In addition, the desire to avoid the
assumption of special initial conditions implied by the FRW models suggests
the study of more general cosmological models such as anisotropic and
spatially homogeneous ones.

It is generally accepted that the early universe was characterized by a highly
irregular expansion mechanism. Therefore, it is interesting to study more
appropriate cosmological models in which anisotropies, existing at early stage
of the expansion, are damped out in the course of the evolution. This
investigation has increased since in Ref. \cite{Hu0} it was shown that the
creation of scalar particles can dissipate the anisotropy as the Universe
expands. A physically acceptable cosmology should provide a mechanism to
achieve an accelerating expansion at the present time.

A Bianchi type-I universe is of particular interest because it is one of the
simplest models that contains special isotropic cases and allows arbitrarily
small anisotropic levels \cite{ja}. Since the actual Universe is surprisingly
isotropic, it makes a suitable candidate for studying the possible effects of
an anisotropy in the early Universe on present-day observations.

The solutions of the low-energy string cosmological effective action are by
their nature anisotropic \cite{pbb} and there is evidence that the dynamics of
the early universe may have been profoundly influenced by the presence of
spatial anisotropies just below the Planck or string scale \cite{1}. In
\cite{mel} a new family of exact spherically symmetric solutions was found in
the model with a one-component anisotropic fluid for a general equation of
state. In this context, it is important to analyze the role of different
sources of anisotropy in the Einstein equations.

Internal symmetry has proven to be a useful tool for field theory in flat
spacetime to understand charge conservation, isospin, etc., and gauge theories
have played an important role in understanding high-energy phenomena. Any
theory that attempts to unify the different physical interactions must discuss
the two following points: the structure of the dynamical interactions among
the fundamental fields, and the nature of the internal symmetry group (ISG)
that governs these interactions. We are interested in the second point, not
with a phenomenological motivation, but for a more fundamental reason, e.g.,
the properties of the Einstein gravitational action itself. As far as we know,
few efforts have been made to investigate this kind of symmetry in the context
of general relativity.

In \cite{sif} we exploited and investigated an unusual kind of internal
symmetry of the Einstein field equations in homogeneous spacetime, e.g., the
form invariance symmetry, which relates the expansion rate ({\it geometrical
variable}) with the energy density and pressure of the perfect fluid ({\it
source variables}). It has proved to be a useful concept to give a
satisfactory explanation of the assisted inflation mechanism, appealing to the
cooperative effect of adding the energy density into the Friedmann equation.

Below, we investigate the internal symmetries of the Einstein equations for
the particular case of an expanding universe described by anisotropic
Bianchi type-I metrics throughout its whole evolution.

The paper is organized as follows. In Sec. II we find the linear
transformation of the expansion rates which preserves the Einstein
gravitational action for the Bianchi type-I spacetime and present the
three-dimensional internal space. In Sec. III we show that a vector
representation of the ISG exists where two scale factors of the Bianchi type-I
cosmology with perfect fluid become equal. We find the general solution for a
perfect fluid with a constant baryotropic index and investigate its asymptotic
behavior. In Sec. IV exact solutions are found for three types of anisotropic
fluids.  Finally, the conclusions are stated in Sec. V.

\section{Internal symmetry group}

The spatially homogeneous anisotropic Bianchi type-I spacetime is described
by the line element

\begin{equation}
\n{b1}
ds ^2=-dt^2+a_1(t)^2dx^2+a_2(t)^2dy^2+a_3(t)^2dz^2.
\end{equation}

\no We write the classical Einstein gravitational action $S_E$ for the metric
(\ref{b1}),

$$
S_E[g]=\int\sqrt{-g}\left[\dot H_1+\dot H_2+\dot H_3+H_1^2+H_2^2+H_3^2+\right.
$$
\be
\label{se}
\left. H_1H_2+H_1H_3+H_2H_3\right]\,d^4x,
\ee

\noindent where $g$ denotes the metric determinant, $H_i=\dot a_i/a_i$ are the
expansion rates along the three spatial directions, and $i=1,2,3$. Throughout
this paper, we use units such that $c=8\pi G=1$. This gravitational action is
invariant under a linear transformation of the expansion rates that preserves
the following quantities:

\be
\n{q1}
Q_1=H_1+H_2+H_3,
\ee

\be
\n{q2}
Q_2=H_1^2+H_2^2+H_3^2.
\ee

\no These invariants can be expressed in terms of the expansion scalar
$\theta$ and the shear scalar $\sigma$ as $Q_1=\theta$ and
$Q_2=\sigma^2+\theta^2/3$. Introducing an internal three-dimensional cartesian
space, we can see that the points of the circumference defined by the
intersection of the plane (\ref{q1}) with the surface of the sphere (\ref{q2})
are equivalent. Hence, the linear transformations $H'_i=H'_i(H_i)$, which
transform points of this circumference into points of the same circumference,
are internal symmetries of the Bianchi type-I geometrical action. A matrix
representation of the ISG, isomorphic to the rotation group on a plane, is
given by

\begin{eqnarray}
{\bf M}=\frac{1}{3}
\left(\begin{array}{ccc}
1+2\cos{\phi} & 1-\cos{\phi}-\sqrt{3}\sin{\phi} & 1-\cos{\phi}+
\sqrt{3}\sin{\phi}\\
1-\cos{\phi}+\sqrt{3}\sin{\phi} & 1+2\cos{\phi}  & 1-\cos{\phi}-
\sqrt{3}\sin{\phi}\\
1-\cos{\phi}-\sqrt{3}\sin{\phi} & 1-\cos{\phi}+\sqrt{3}\sin{\phi} &
1+2\cos{\phi}\\
\end{array}\right),
\n{M}
\end{eqnarray}

\no where $\phi$ is the rotation angle. The elements of the rotation matrix
${\bf M}$ have the additional properties $\sum^3_{i=1} {\bf M}_{ik}=1$ and
$\sum^3_{k=1} {\bf M}_{ik}=1$ to preserve the quantity $Q_1$ defined in
Eq.  (\ref{q1}).

The variational problem in the gravitational action (\ref{se}), for a
stress tensor representing a fluid with energy density $\rho$ and pressures
$p_i$, leads to the Einstein equations

\be
\n{00}
3H^2=\rho+\frac{1}{2}\sigma^2,
\ee

\be
\n{11}
-2\dot H=p+\rho+\sigma^2,
\ee

\be
\n{cf}
\dot\rho+3H(\rho+p)+\vec\sigma\cdot\vec\Sigma=0,
\ee

\be
\n{cs}
\vec\sigma\cdot\dot{\vec\sigma}+3H\vec\sigma^2=\vec\sigma\cdot\vec\Sigma.
\ee

\no where the average expansion rate $H$ and the shear vector $\vec\sigma$ are
the trace and the traceless part of the extrinsic curvature $K_{ij}$ of the
homogeneous time slice. The former can be expressed in terms of the average
scale factor $a= (-g)^{1/6}$, as $H=\dot a/a$, and the latter has components
$\sigma_i=H_i-H$. The isotropic pressure is $p= ( p_1+p_2+p_3 )/3$ and the
components of the vector $\vec\Sigma$ are the transverse pressures
$\Sigma_i=p_i-p$. In addition, $\vec\sigma$ and $\vec\Sigma$ transform as a vector
in the internal space. Note that decomposing the three-dimensional internal
space as a direct sum of one-dimensional space containing the rotation axis
and a two-dimensional orthogonal space, we can see that $H$ and $p$ are
invariants under the action of the ISG, while $\vec\sigma$ and $\vec\Sigma$
are vectors contained in the two-dimensional orthogonal space.

From the dynamical equation of the shear (\ref{cs}) we get

\be
\n{s}
\sigma_i=\frac{\sigma_{i0}}{a^3}+\frac{1}{a^3}\int a^3\Sigma_i\,dt,
\ee

\no where the three integration constants $\sigma_{i0}$ transform as the
components of a vector in the internal space and satisfy the transverse
condition $\sigma_{10}+\sigma_{20}+\sigma_{30}=0$ with
$\sigma^2_{10}+\sigma^2_{20}+\sigma^2_{30}=\sigma_0^2$. Defining the
dimensionless constants $q_i=\sigma_{i0}q/\sigma_0$ such that

\be
\n{cq}
q_1+q_2+q_3=0, \qquad q^2_1+q^2_2+q^2_3=q^2,
\ee

\no along with the functions $m$ and $\mu_i$,
\be
\n{mm}
\frac{\dot m}{m}=\left(\frac{a_0}{a}\right)^{3},  \qquad
\frac{\dot\mu_i}{\mu_i}=\frac{1}{a^3}\int a^3\Sigma_i\,dt,
\ee

\no where $a_0=(\sigma_0/q)^{1/3}$, then the scale factors along the three
spatial directions are obtained by integrating the shear vector (\ref{s})

\be
\n{A}
a_i=a_{i0}\,a\,\mu_i\,m^{q_i},
\ee

\no where the integration constants $a_{i0}$ and the functions $\mu_i$ satisfy
the constraint $a_{10}a_{20}a_{30}\mu_1\mu_2\mu_3=1$.

Due to the vector nature of the expansion rates $H_i$ the three scale factors
$a_i$ give rise to the following vector representation of the ISG

\be
\n{ta}
\ln a'_i={\bf M}_{ik}\ln a_k,
\ee

\no and the internal symmetry transformations map solutions of the Einstein
equations (\ref{00})-(\ref{cs}) into solutions of the same equations. In the
next section we will use Eq. (\ref{ta}) to reduce the internal degrees of
freedom of the Bianchi type-I metric. On the other hand, the transformation
rule (\ref{ta}) extends the results of the previous paper \cite{sif}, where
the form invariance symmetry of the Einstein equations for a
Friedmann-Robertson-Walker spacetime was shown .

\section{Bianchi type-I cosmology with a perfect fluid}

A wide class of cosmological models have the initial stage similar to that
of the vacuum Bianchi type-I model first considered by Kasner. Investigations
of such spacetime filled with a perfect fluid were carried out in
\cite{he}-\cite{bar}. Later on in \cite{ka},\cite{bar} it was shown that
the evolution of the model with a stiff fluid is significantly different from
that of models with dust or radiation, and the initial vacuum assumption can
be violated.

In this section we use the internal space to reexamine the Bianchi type-I
cosmology with a perfect fluid characterized by an isotropic pressure $p$. For
this fluid the transverse pressure $\vec\Sigma$ vanishes, so Eq. (\ref{mm})
gives $\mu_i=$const and the shear (\ref{s}) turns into

\be
\n{qi}
\sigma_i=\frac{\sigma_{i0}}{a^3}.
\ee

\no Then $\sigma^2=\sigma^2_{0}/a^6$ is the contribution of the shear to Eq.
(\ref{00}) and the Einstein equations (\ref{00})-(\ref{cs}) look as if they
were the Friedmann-Robertson-Walker equations for two separately conserved
perfect fluids, one of which behaves as if it were a stiff fluid. In the
$q_{i}$-parameter representation the scale factors (\ref{A}) read

\be
\n{Ai}
a_i=a_{i0}\,a\,m^{q_i},
\ee

\no where the three integration constants $a_{i0}$ satisfy the condition
$a_{10}a_{20}a_{30}=1$.

As Eqs. (\ref{00})-(\ref{cs}) contain an explicit dependence on the
constant $q^2$ and each scale factor (\ref{Ai}) is generated by the constants
$q_i$, two different solutions with labels $q_i$ and $q'_i$ satisfying
Eq. (\ref{cq}) can be related by an internal symmetry transformation. The
linear transformation $q'_i~=~{\bf M}_{ik}q_k$ induces the transformation
(\ref{ta}) between two different solutions of the Einstein equations with the
same source and shear scalar.

The ISG can be used to reduce the internal degrees of freedom of the Bianchi
type-I metric. In fact, the scale factors $a_i$ generate a vector
representation of this group; hence, for a given set of constants $q_i$ a
system of coordinates exists in the internal space where two components of the
transformed constants $q'_i$ become equal. This job is done by the matrix

\begin{eqnarray}
\n{M1}
{\bf M}_{(q'_1=q'_2)}=\frac{1}{3}
\left(\begin{array}{ccc}
1-\sqrt{6}\,q_3/q & 1-\sqrt{6}\,q_1/q & 1-\sqrt{6}\,q_2/q\\
1-\sqrt{6}\,q_2/q & 1-\sqrt{6}\,q_3/q & 1-\sqrt{6}\,q_1/q\\
1-\sqrt{6}\,q_1/q & 1-\sqrt{6}\,q_2/q & 1-\sqrt{6}\,q_3/q\\
\end{array}\right).
\end{eqnarray}

\no Using this matrix in the transformation rule (\ref{ta}), we obtain the
scale factors in the new representation

\be
\n{tai}
a'_1=a_{10}\,a\,m^{q/\sqrt{6}},
\qquad a'_2=a_{20}\,a\,m^{q/\sqrt{6}},
\qquad a'_3=a_{30}\,a\,m^{-\sqrt{2/3}\,q}.
\ee

\no This shows that any Bianchi type-I cosmology with a perfect fluid can be
mapped by means of an internal symmetry transformation to a locally
rotationally symmetric model with the same source and shear scalar. To see
this result explicitly we fix the constant $q^2=2/3$ in the first Eq.
(\ref{mm}); then the scale factors (\ref{Ai}) become

\be
\n{ai}
a_1=a_{10}a_0\,\frac{m^{s_1}}{\dot m^{1/3}}, \qquad
a_2=a_{20}a_0\,\frac{m^{s_2}}{\dot m^{1/3}},   \qquad
a_3=a_{30}a_0\,\frac{m^{s_3}}{\dot m^{1/3}},
\ee

\no where the parameters $s_i=q_i+1/3$ satisfy the Kasner constraints

\be
\n{ps}
s_1+s_2+s_3=1,	\qquad	s^2_1+s^2_2+s^2_3=1.
\ee

\no A one-parameter representation of the Kasner exponents $s_i$ is given by

\be
\n{pb}
s_1=\frac{1}{3}\left[1+\frac{2b}{\sqrt{3+b^2}}\right], \,
s_2=\frac{1}{3}\left[1+\frac{3-b}{\sqrt{3+b^2}}\right], \,
s_3=\frac{1}{3}\left[1-\frac{3+b}{\sqrt{3+b^2}}\right].
\ee

\no where $b$ is the parameter. Hence, for the Bianchi type-I spacetime filled
with a perfect fluid, we conclude that the set of solutions (\ref{ai})
generated by the Kasner exponents $s_i$ are related to each other by internal
symmetry transformations. In particular, for $b=1$ we reobtain the locally
rotationally symmetric solution (\ref{tai}).

\subsection{General solution}

Now, we investigate a cosmological model with perfect fluid. The
energy-momentum tensor of the fluid is $T_{ik}=\rho u_{i}u_{k}+ph_{ik}$, where
$\rho$ is the energy density, $p$ the equilibrium pressure, $u^i$ the
four-velocity of the fluid, and $h_{ik}$ the projection tensor
$h_{ik}=g_{ik}+u_{i}u_{k}$. The components of the fluid four-velocity are
$u^i=(1,0,0,0)$. An equation of state of the form $p=(\gamma-1)\rho$ with a
constant baryotropic index $\gamma$ is also assumed. Integrating the
conservation Eq. (\ref{cf}) we get the energy density
$\rho=\Lambda/a^{3\gamma}$, where $\Lambda$ is an integration constant. In
this case, introducing the variable $v=a^3=\sqrt{-g}$ in Eqs. (\ref{00})
and (\ref{mm}), they become

\be
\n{v}
v'^2=1+3\lambda v^{2-\gamma}, \qquad \frac{m'}{m}=\frac{1}{v},
\ee

\no where the prime indicates differentiation with respect to the
dimensionless time $T=\sqrt{3/2}\,\sigma_0t$ and
$\lambda=2\Lambda/3\sigma_0^2$. The general solutions of both equations in
(\ref{v}) are given as follows.

\vskip 1cm

\no {\it Case $\lambda>0$ (positive energy density)}

\be
\n{vg}
a^3=\left[\frac{1}{3\lambda}\sinh^2{(\tau-\tau_0)}\right]^{1/(2-\gamma)},
\qquad m=\left[\tanh{\frac{(\tau-\tau_0)}{2}}\right]^{2/(2-\gamma)},
\ee

\be
\n{t}
T=\frac{2}{2-\gamma}(3\lambda)^{1/(\gamma-2)}
\int \left[\sinh\tau\right]^{\gamma/(2-\gamma)}\,d\tau\,,
\ee

\no where $\tau_0$ is an integration constant. From Eqs. (\ref{ai}),
(\ref{vg}) and (\ref{t}) the scale factors are

\be
\n{aig}
a_i=a_{i0}\left(\frac{4}{3\lambda}\right)^{1/3(2-\gamma)}
\left[\cosh{\frac{(\tau-\tau_0)}{2}}\right]^{4/3(2-\gamma)}
\left[\tanh{\frac{(\tau-\tau_0)}{2}}\right]^{2s_i/(2-\gamma)}.
\ee

\no If the baryotropic index is restricted to the range $0<\gamma<2$, then
from Eq. (\ref{t}) we see that $T$ and $\tau$ have the same asymptotic limits.
This allows us to investigate the evolution of the scale factors in two
asymptotic regimes using the exact solutions (\ref{vg})-(\ref{aig}). In the
first regime, $a^3<(3\lambda)^{1/(\gamma-2)}$, the perfect fluid is
dynamically unimportant, the approximate average scale factor (\ref{vg})
becomes $a^3\propto T$, and $m\propto T$. Hence, qualitatively the behavior of
the scale factors (\ref{aig}) is similar to that of the vacuum Kasner
solution. Then, in first approximation, the fluid can be considered as a test
matter on a highly anisotropic background. In the second regime, which starts
from some characteristic time where $a^3>(3\lambda)^{1/(\gamma-2)}$, the fluid
becomes dominant. In this case, the approximate average scale factor
(\ref{vg}) and the scale factors (\ref{aig}) are $a\propto a_i\propto
T^{2/3\gamma}$. This result shows that, owing to the spatial isotropy of the
stress-energy tensor, the anisotropic Bianchi type-I model evolves into a
Friedmann-Robertson-Walker cosmology and the initial anisotropy of this simple
cosmological model is dissipated as the Universe expands.

\vskip 1cm

\no {\it Case $\lambda<0$ (negative energy density)}

\be
\n{vg-}
a^3=\left[-\frac{1}{3\lambda}\sin^2{(\tau-\tau_0)}\right]^{1/(2-\gamma)},
\qquad m=\left[\tan{\frac{(\tau-\tau_0)}{2}}\right]^{2/(2-\gamma)},
\ee

\be
\n{t-}
T=\frac{2}{2-\gamma}(-3\lambda)^{1/(\gamma-2)}
\int \left[\sin\tau\right]^{\gamma/(2-\gamma)}\,d\tau\,.
\ee

\no From Eqs. (\ref{ai}), (\ref{vg-}) and (\ref{t-}) the scale factors
are

\be
\n{ag-}
a_i=a_{i0}\left(-\frac{4}{3\lambda}\right)^{1/3(2-\gamma)}
\left[\cos{\frac{(\tau-\tau_0)}{2}}\right]^{4/3(2-\gamma)}
\left[\tan{\frac{(\tau-\tau_0)}{2}}\right]^{2s_i/(2-\gamma)},
\ee

\no they represent a universe having a finite time span.

For the particular cases $\lambda=0$ and $\gamma=2,1,0$, the explicit general
solutions are given below.

\vskip 1cm

\no {\it 1.} $\lambda=0$ (vacuum),

\be
\n{0}
a^3=T, \qquad  m=T, \qquad  a_i=a_{i0}\,T^{s_i}.
\ee

\vskip 1cm

\no {\it 2.} $\gamma=2$ (stiff fluid),

$$
a^3=\sqrt{1+3\lambda}\, T,      \qquad  m=T^{1/\sqrt{1+3\lambda}},
$$

\be
\n{2}
a_i=a_{i0}\,(1+3\lambda)^{1/6}\,T^{\frac{1}{3}+
\frac{(s_i-1/3)}{\sqrt{1+3\lambda}}}.
\ee

\vskip 1cm

\no {\it 3.} $\gamma=1$ (dust),

$$
a^3=T\left(1+\frac{3\lambda}{4}T\right),
\qquad m=\frac{T}{T+\frac{4}{3\lambda}},
$$

\be
\n{12}
a_i=a_{i0}\left(\frac{3\lambda}{4}\right)^{1/3}
\left[\frac{T}{T+\frac{4}{3\lambda}}\,\right]^{s_i}
\left(T+\frac{4}{3\lambda}\right)^{2/3}.
\ee

\vskip 1cm

\no {\it 4.} $\gamma=0$ (positive cosmological constant),

$$
a^3=\frac{1}{\sqrt{3\lambda}}\sinh{\sqrt{3\lambda}\,T},  \qquad
m=\tanh{\frac{\sqrt{3\lambda}}{2}\,T},
$$

\be
\n{0+}
a_i=a_{i0}\left(\frac{2}{\sqrt{3\lambda}}\right)^{1/3}\left[
\tanh{\frac{\sqrt{3\lambda}}{2}\,T}\right]^{s_i}
\cosh^{2/3}{\frac{\sqrt{3\lambda}}{2}\,T}.
\ee

\vskip 1cm

\no {\it 5.} $\gamma=0$ (negative cosmological constant),

$$
a^3=\frac{1}{\sqrt{-3\lambda}}\sin{\sqrt{-3\lambda}\,T},  \qquad
m=\tan{\frac{\sqrt{-3\lambda}}{2}\,T},
$$

\be
\n{0-}
a_i=a_{i0}\left(\frac{2}{\sqrt{-3\lambda}}\right)^{1/3}\left[
\tan{\frac{\sqrt{-3\lambda}}{2}\,T}\right]^{s_i}
\cos^{2/3}{\frac{\sqrt{-3\lambda}}{2}\,T}.
\ee

Explicit solutions can also be found for other $\gamma$ values, but for
physical reasons we do not pay attention to them.

\section{Anisotropic fluid}

The kinetic and potential energy of the gravitational field that may be
attributed to the presence of anisotropy could reach the same order of
magnitude or even exceed the energy attributed to any other source present in
the effective action at some time in the early evolution. Hence, the dynamics
of the universe at that time could be profoundly distorted if one takes into
account that contribution. To investigate these effects it is convenient to
decompose the vector pressure in the internal space  as a sum of a vector
along the rotation axis vector $\vec n=(1,1,1)$ and a vector contained in a
plane perpendicular to this axis, then $\vec{\cal P}=p\,\vec n+\vec\Sigma$.

Now we can find exact solutions of the Einstein equations
(\ref{00})-(\ref{cs}) for three different sources,  an anisotropic fluid with
constant stress-energy tensor, an anisotropic fluid characterized by a vector
pressure whose components are proportional to the energy density, and an
anisotropic stiff fluid.

\subsection{Constant stress-energy tensor}

This fluid has a constant energy density $\rho$ and a constant vector pressure
$\vec{\cal P}~=~(p_1,p_2,p_3)$. From Eqs. (\ref{00}), (\ref{11}) and
(\ref{s}), we have

\be
\n{vc}
(a^3)^{\cdot\cdot}-\beta a^3=0,   \qquad    \beta=\frac{3}{2}(\rho-p),
\ee

\no and

\be
\n{ss}
\sigma_i=\frac{\sigma_{i0}}{a^3}+3\frac{\Sigma_i}{\beta}H,
\ee

\no where $\Sigma_i=p_i-p$ are the components of the transverse pressure. On
the other hand, the conservation Eq. (\ref{cf}) gives

\be
\n{ee}
\rho+p+\frac{\vec\sigma_{0}\cdot\vec\Sigma}{3Ha^3}+\frac{\Sigma^2}{\beta}=0,
\qquad \Sigma^2=\vec\Sigma.
\ee

\no Choosing the integration constants so that the vectors $\vec \sigma_{0}$
and $\vec\Sigma$ are perpendicular, the third term in Eq. (\ref{ee})
vanishes. This choice, which is invariant under the action of the ISG, selects
solutions for which the equation of state is

\be
\n{e}
p^2=\rho^2+\frac{2}{3}\Sigma^2,
\ee

\no or in terms of the pressures $p_i$

\be
\n{epi}
2\left(p_1p_2+p_1p_2+p_2p_3\right)=p_1^2+p_2^2+p_3^2+3\rho^2.
\ee

\no When the energy density $\rho$ is greater than the transverse pressure
$\Sigma$, we expand the equation of state (\ref{e}) in powers of $\Sigma/\rho$
and $p\approx\pm(\rho+\Sigma^2/3\rho+...)$. The negative branch could describe
a modified Chaplygin gas (see  \cite{Be}); by a Chaplygin gas we mean a
perfect fluid with the equation of state $p=-A/\rho$ where $A$ is a constant
and $p < 0$. This equation of state has raised a certain interest recently due
to its many interesting features \cite{Ba}. It has a connection with string
theory and can be obtained from the Nambu-Goto action for $d$-branes moving in
a $(d+2)$-dimensional spacetime in the light-cone parametrization \cite{Bor}.
In addition, the Chaplygin gas is the only fluid which, up to now, admits a
supersymmetric generalization \cite{Ho,Ja}. In the above approximation the
constant $A=\Sigma^2/3$ would represent the first contribution of the
anisotropy pressure to the isotropic average pressure.

From Eqs. (\ref{mm}) and (\ref{ss}) we obtain the scale factors

\be
\n{sd}
a_i=a_{i0}\,m^{q_i}\,a^{1+\frac{3\Sigma_i}{\beta}},
\ee

\no where the variable $v=a^3$ satisfies the equation

\be
\n{00d}
\dot v^2=\frac{3}{4}\sigma_0^2\,\frac{\rho-p}{\rho}\left[1+
\frac{2\rho}{\sigma_0^2}v^2\right],
\ee

\no which has been obtained by inserting Eq. (\ref{ss}) in Eq. (\ref{00}). For
$\rho=-p=\lambda$ Eq. (\ref{00d}) reduces to Eq. (\ref{v}) for $\gamma=0$, Eq.
(\ref{e}) leads to a vanishing transverse pressure, and the fluid becomes a
perfect fluid. In other cases, introducing the parameter
$3\lambda=2\rho/\sigma_0^2$ and the dimensionless time
$T=\sigma_0\sqrt{3(\rho-p)/\rho}\,\,t/2$ in Eqs. (\ref{mm}) and (\ref{00d}),
they turn into Eq. (\ref{v}) for $\gamma=0$ after fixing
$q=\sqrt{2\rho/\beta}$. Inserting its solutions in the scale factors
(\ref{sd}) we get three different types of solutions

\vskip 1cm

\no {\it 1.} $\rho>0$ and $p<-\rho$

$$
a^3=\frac{\sigma_0}{\sqrt{2\rho}}\sinh{\sqrt{\beta}\,t},  \qquad
m=\left[\tanh{\frac{\sqrt{\beta}}{2}\,t}\right],
$$
\be
\n{1d}
a_i=a_{i0}\left[\tanh{\frac{\sqrt{\beta}}
{2}\,t}\right]^{q_i}\left(\frac{\sigma_0}{\sqrt{2\rho}}
\sinh{\sqrt{\beta}\,t}\right)^{\frac{1}{3}+\frac{\Sigma_i}{\beta}}.
\ee

\vskip 1cm

\no {\it 2.} $\rho<0$ and $p>-\rho$

$$
a^3=\frac{\sigma_0}{\sqrt{-2\rho}}\sin{\sqrt{-\beta}\,t},  \qquad
m=\left[\tan{\frac{\sqrt{-\beta}}{2}\,t}\right],
$$
\be
\n{2d}
a_i=a_{i0}\left[\tan{\frac{\sqrt{-\beta}}
{2}\,t}\right]^{q_i}\left(\frac{\sigma_0}{\sqrt{-2\rho}}
\sin{\sqrt{-\beta}\,t}\right)^{\frac{1}{3}+\frac{\Sigma_i}{\beta}}.
\ee

\vskip 1cm

\no {\it 3.} $\rho<0$ and $p<\rho$

$$
a^3=\frac{\sigma_0}{\sqrt{-2\rho}}\cosh{\sqrt{\beta}\,t},  \qquad
m=\exp\left[2\tan^{-1}{{\mbox e}^{\sqrt{\beta}\,t}}\right],
$$
\be
\n{3d}
a_i=a_{i0}\exp\left[2q_i\tan^{-1}{\mbox e}^{\sqrt{\beta}\,t}\right]
\left(\frac{\sigma_0}{\sqrt{-2\rho}}
\cosh{\sqrt{\beta}\,t}\right)^{\frac{1}{3}+\frac{\Sigma_i}{\beta}}.
\ee

Solutions (\ref{1d}) and (\ref{2d}) could be considered as a generalization of
the solutions (\ref{0+}) and (\ref{0-}) with constant cosmological term. In
fact, when the transverse pressure vanishes both sets of solutions are the
same.

\subsection{$p_i\propto\rho$}

Here we will find exact solutions for a cosmological Bianchi type-I model with
an anisotropic fluid characterized by a vector pressure whose components are
proportional to the energy density and investigate their asymptotic behavior.
The equation of state is

\be
\n{pa}
\vec{\cal P}=(\vec{\cal G}-\vec n)\rho,
\ee

\no with $\vec{\cal G}=(\gamma_1,\gamma_2,\gamma_3)$ a constant vector index
that can be written as

\be
\n{g}
\vec{\cal G}=\vec{\cal G}^n+\vec{\cal G}^\pi=\gamma\,\vec n+\vec\Gamma,
\ee

\no where $0\le\gamma<2$ is the average index and $\vec\Gamma$ is the
transverse index. The particular case of $\gamma=2$, anisotropic stiff fluid,
will be investigated in the next subsection. Comparing Eqs. (\ref{pa}) and
(\ref{g}) with $\vec{\cal P}=p\,\vec n+\vec\Sigma$, the average pressure and
the transverse pressure are

\be
\n{pat}
p=(\gamma-1)\rho,  \qquad   \vec\Sigma=\vec\Gamma\rho.
\ee

\no For vanishing transverse index the average pressure $p$ reduces to the
isotropic pressure and the equation of state becomes that of a perfect fluid.

Combining Eqs. (\ref{00})-(\ref{11}) with the first one in Eq.
(\ref{pat}), we obtain

\be
\n{v..}
a^3\rho=\frac{2}{3(2-\gamma)}(a^3)^{\cdot\cdot}\,.
\ee

\no Using Eqs. (\ref{s}), (\ref{mm}),  (\ref{pat}) and (\ref{v..}), we
calculate the shear vector

\be
\n{sp}
\vec\sigma=\frac{\vec\sigma_0}{a^3}+\frac{2\vec\Gamma}{2-\gamma}H,
\ee

\no and the corresponding scale factors after integrating Eq. (\ref{sp})

\be
\n{aia}
a_i=a_{i0}\,m^{q_i}a^{1+\frac{2\Gamma_i}{2-\gamma}}.
\ee

With the aid of Eq. (\ref{mm}), the transverse pressure (\ref{pat}),
and the shear vector (\ref{sp}) we can integrate the conservation Eq.
(\ref{cf}) to obtain the energy density of the anisotropic fluid

\be
\n{rp}
\rho=\frac{\Lambda\,m^{-q\frac{\vec\sigma_0\cdot\vec\Gamma}{\sigma_0}}}
{a^{3\gamma+\frac{2\Gamma^2}{2-\gamma}}},
\ee

\no where $\Lambda$ is an integration constant and
$\Gamma^2=\gamma_1^2+\gamma_2^2+\gamma_3^2-3\gamma^2$. Defining the effective
index $\gamma_e=\gamma+\frac{2\Gamma^2}{3(2-\gamma)}$ and choosing the vector
$\vec\sigma_0$ perpendicular to $\vec\Gamma$, the energy density of the fluid
takes the form $\rho=\Lambda/a^{3\gamma_e}$ and the Einstein equation
(\ref{00}) for the variable $v=a^3$ becomes

\be
\n{v.p}
\dot v^2=\frac{3}{2}\sigma_0^2\frac{2-\gamma}{2-\gamma_e}\left[1+
\frac{2\Lambda}{\sigma_0^2}v^{2-\gamma_e}\right].
\ee

\no Finally, the scale factors are obtained after inserting the solutions of
Eqs. (\ref{mm}) and (\ref{v.p}) in Eq. (\ref{aia}). For fluids with
average index $0\le\gamma<2$, the contribution of the transverse pressure
increases the effective index. Consequently, the anisotropic pressure
decelerates the average expansion of the universe.

The use of the dimensionless time
$T=\sigma_0\sqrt{3(2-\gamma)/2(2-\gamma_e)}\,\,t$ transforms Eqs. (\ref{mm})
and (\ref{v.p}) into Eq. (\ref{v}) for $3\lambda=2\Lambda/\sigma_0^2$ and
$\gamma=\gamma_e$ after fixing $q=\sqrt{2(2-\gamma_e)/3(2-\gamma)}$. Its
implicit solutions are obtained making these changes of variables in the
solutions (\ref{vg}),(\ref{t}) and (\ref{vg-}),(\ref{t-}) of the Sec. {\bf III
A.}. Explicit solutions also can be found for $\gamma_e=0,1,2$ following the
same steps as were done in that subsection. For a vanishing transverse
pressure $\vec\Sigma$ we reobtain Eq. (\ref{v}) for the isotropic perfect
fluid.

To look into anisotropic fluid effects on the cosmological model, the
behavior of the scale factors (\ref{aia}) will be investigated in two
asymptotic regimes assuming that $0<\gamma<2$ and $\gamma_e<2$. In the first
regime, when $a^3 <|\sigma^2_0/2\Lambda|^{1/(2-\gamma_e)}$ the contribution of
the anisotropic fluid can be neglected and Eqs. (\ref{v.p}), (\ref{mm})
give $v\approx T$ and $m\approx T$. Hence, the scale factors (\ref{aia})
behave as

\be
\n{an}
a_i\approx a_{i0}\,T^{\frac{1}{3}+q_i+\frac{2\Gamma_i}
{3(2-\gamma)}}.
\ee

\no In the second regime, which starts when
$a^3>|\sigma^2_0/2\Lambda|^{1/(2-\gamma_e)}$ the fluid dominates and Eqs.
(\ref{mm}), (\ref{v.p}) have the approximate solutions $a^3\approx \nu
T^{2/\gamma_e}$ and $m\approx
$exp$[(\gamma_e/\nu(\gamma_e-2))T^{(\gamma_e-2)/\gamma_e}]$, where
$\nu=(\gamma_e\sqrt{3\lambda}/2)^{2/\gamma_e}$ is a constant. Inserting them
in Eq. (\ref{aia}), we get the large time behavior of the scale factors

\be
\n{ani}
a_i\approx a_{i0}\left[\nu^{1/3} T^{2/3\gamma_e}\right]^
{1+\frac{2\Gamma_i}{2-\gamma}}.
\ee

\no Therefore, this cosmological model never reaches an isotropic stage. For
$\gamma_e>2$ the existence of the solutions is linked to a negative energy
density and the average scale factor has a finite time span.

\subsection{Anisotropic stiff fluid}

In this last subsection we will investigate the particular case of an
anisotropic stiff fluid with equation of state $p=\rho$, where $p$ is the
average pressure, and find the general exact solution for a selected set of
transverse pressures. The average index (\ref{pat}) of this fluid is
$\gamma=2$ and the general solution of Eq. (\ref{vc}) is $a^3=a_0^3t$, where
$a_0$ is a constant. So the average expansion rate is $H=1/3t$. This suggests
analyzing a transverse pressure which depends on the average scale factor as

\be
\n{st}
\vec\Sigma=\frac{\vec\Gamma}{a^{3n}},
\ee

\no where the transverse vector $\vec\Gamma$ and the parameter $n$ are
constants. The shear vector (\ref{s}) becomes

\be
\n{sst}
\vec\sigma=\left(\frac{a_0}{a}\right)^3\,\vec q+
\frac{\vec\Gamma}{(2-n)a_0^3a^{3(n-1)}},
\qquad n\ne 2,
\ee

\be
\n{sst'}
\vec\sigma=\left(\frac{a_0}{a}\right)^3\,\vec q+
\frac{\vec\Gamma}{(a_0a)^3}\ln{a^3},
\qquad n=2,
\ee

\no and Eq. (\ref{00}) gives the relation $\rho=a_0^6/3a^6-\sigma^2/2$
between the energy density of the anisotropic stiff fluid and the shear
scalar.

Integrating the shear vectors (\ref{sst}) and (\ref{sst'}), we get the general
solution

\be
\n{ast}
a_i=a_{i0}\,a_0\,t^{\frac{1}{3}+q_i}\,{\mbox {exp}}\left[\frac{
\Gamma_i t^{(2-n)}}{(2-n)^2\,a_0^{3n}}\right], \qquad n\ne 2,
\ee

\be
\n{ast'}
a_i=a_{i0}\,a_0\,t^{\frac{1}{3}+q_i+\frac{\Gamma_i}{2a_{0}^6}\ln{a_0^3\,t}},
\qquad n=2,
\ee

\no For $n<2$ the scale factors begin to evolve as it were an isotropic stiff
fluid, $a_i\approx t^{\frac{1}{3}+q_i}$, (see Eq. (\ref{2})). But at late time
the anisotropic stiff fluid dominates and, in this regime, the behavior of
the scale factors is quite different from the isotropic case (see Eq.
(\ref{2})). For $n >2$, the model has a high degree of anisotropy at early
time, where the anisotropic stiff fluid dominates. After that, the scale
factors evolve into those given by Eq. (\ref{2}). This model could be
used to describe a universe that is anisotropic at the very beginning. For
$n=2$, the solution behaves as solution (\ref{2}) near the characteristic time
$t_c=1/a_0$.

\section{Conclusions}

We have found in the above results that the ISG of the Bianchi type-I
cosmology is given internally by the Einstein gravitational action rather than
being an additional external hypothesis imposed upon the model. Such a
possibility appears to be reasonable for any unified gauge theory so that (a)
the interactions are completely determined by the gauge invariance, and (b)
{\it a priori}, the theory possesses an internal gauge group.

The relation between this ISG and the geometrical properties of the model is
made through parameters such as the expansion scalar and the shear scalar,
this means that the invariants of the group are expressed in terms of these
quantities. The decomposition of the three-dimensional internal space as a
direct sum of one-dimensional space containing the rotation axis and a
two-dimensional orthogonal space allows one to write the Einstein equations in
a simple manner. The three scale factors generate a vector representation of
the ISG, and the associated transformations include a set that has been used
to reduce the Bianchi type-I cosmology with a perfect fluid to a locally
rotationally symmetric one with the same source and shear scalar. The
existence of this particular representation is guaranteed by the existence of
the internal space itself.

We have given an integral representation of the general solution to the
Bianchi type-I cosmological model with a perfect fluid and constant
baryotropic index, and shown that the scale factors approach the average
scale factor asymptotically for large times when $0\le\gamma<2$. These
solutions begin as Kasner solutions, but after some characteristic time they
have a final Friedmann-Robertson-Walker stage. Hence, the perfect fluid source
dissipates the initial anisotropy of the model.

In the last section we considered a Bianchi type-I cosmological model with
three different types of anisotropic fluid; except for the anisotropic stiff
fluid, the solutions found satisfy the constraint $\vec q\cdot\vec\Sigma=0$.
For an anisotropic fluid with constant stress-energy tensor we found a set of
exact solutions which generalize those with a cosmological constant. When the
magnitude of the transverse pressure is significantly smaller than the energy
density of the fluid, the equation of state has a certain resemblance to that
of an exotic fluid \cite{Be}, called a modified Chaplygin gas equation of
state. In the case where each pressure $p_i$ of the anisotropic fluid is
proportional to the energy density $\rho$, we showed that the transverse
pressure increases the effective index and decelerates the average expansion
of the universe. Finally, we found a general solution for an anisotropic stiff
fluid with transverse pressure $\vec\Sigma=\vec\Gamma/a^{3n}$. For $n>2$, the
solutions are highly anisotropic at early time, but at late times they evolve
into the isotropic stiff fluid solutions.

\begin{acknowledgments}

This work was supported by the University of Buenos Aires under Project X223.
LPC thanks Alejandro S. Jakubi for useful discussions of this work.

\end{acknowledgments}


\end{document}